\documentclass[12pt]{iopart} 

\usepackage[utf8]{inputenc}
\usepackage{amssymb}
\usepackage{graphicx}
\usepackage{color}
\usepackage{dcolumn}
\usepackage{bm}
\usepackage{epstopdf}
\usepackage{alltt}

\usepackage[compress]{cite}

\begin{document}

\title{Arch-based configurations in the volume ensemble of static granular systems}

\author{D. Slobinsky$^{1,2}$, Luis A. Pugnaloni$^{1,2}$}

\address{$^1$ Departamento de Ingenier\'ia Mec\'anica, Facultad Regional La Plata, Universidad Tecnol\'ogica Nacional, Avenida 60 Esq. 124, 1900 La Plata, Argentina.\\
$^2$ Consejo Nacional de Investigaciones Cient\'ificas y T\'ecnicas (CONICET), Argentina.}
\ead{luis.pugnaloni@frlp.utn.edu.ar (L A Pugnaloni)}

\begin{abstract}
{We propose an alternative approach to count the microscopic static configurations of granular packs under gravity by considering arches. This strategy simplifies the problem of filtering out configurations that are not mechanically stable, opening the way for a range of granular models to be studied via ensemble theory. Following this arch-based approach, we have obtained the exact density of states for a two-dimensional non-interacting rigid arch model of granular assemblies. The calculated arch size distribution and volume fluctuations show qualitative agreement with realistic simulations of tapped granular beds. We have also validated our calculations by comparing with the analytic solution for the limiting case of a quasi-one-dimensional column of frictionless disks.}
\end{abstract}

\maketitle

\section{Introduction}
Granular materials consist of large collections of particles (grains) that obey the well established macroscopic laws of motion and that interact through a combination of conservative and dissipative forces. Under gravity, in the absence of other external drivings, grains will settle to form a pack. These static packings (generally filling a container but also forming piles on a surface) can be prepared following protocols that warrant statistically reproducible states. An example of these protocols is annealing by tapping \cite{nowak1997reversibility}. The macroscopic steady state obtained by tapping with a given external pulse is defined as the collection of static configurations (the mechanically stable microstates or simply configurations) obtained tap after tap, once any history dependent transient has passed. 

The mechanically stable configurations, are assumed to be amenable of statistical description as pointed out by Edwards \cite{Edwards19891080}. Each of these microstate is fully described by the position of the grains (and their orientations for non-spherical particles) and the force on each contact. The entropy $S(V,\Sigma,N)$ is defined as the logarithm of the number of mechanically stable configurations compatible with total volume $V$, total force moment tensor $\Sigma$ and a number $N$ of grains \cite{edwards2005full,blumenfeld2009granular,C2SM06898B,henkes2007entropy}. Here, each configuration is taken as equi-probable as a basic postulate \cite{Edwards19891080}. It is important to emphasize that the volume and force degrees of freedom are not decoupled, hence the entropy is not the plain sum of the entropies of separated sets of configurations \cite{blumenfeld2012interdependence}. However, in this work we will only consider the pure ``volume ensemble'', associated to the set of geometrical 
configurations that 
the grains can be arranged in without caring for the actual forces needed to equilibrate them in their positions. The majority of contributions in the area have in fact focused in this partial description with fewer devoted to the pure ``force ensemble'' (equally partial) where variations in the particle positions are not included in the description. There is a lack of studies where the states in the full ($V, \Sigma, N$) ensemble are sampled for a particular model.  

The static configurations are ``needles in a haystack'' since the set of mechanically stable microstates has zero measure in the set of all arbitrary particle positions. The traditional approach to deal with this problem relies on filtering out the microstates that are not mechanically stable \cite{review} from the overwhelming set of configurations of the system. This approach is computationally unfeasible except for very small systems \cite{C2SM06898B}. In recent years, some computational techniques have been developed to count all possible static configurations in some systems of sizes of about 7 \cite{gao2009experimental}, 16 \cite{hoy2012structure}, 20 \cite{PhysRevLett.101.128001}, and 128 \cite{asenjo2014numerical} particles. Even if one such static configuration is known, there exists no algorithm capable of performing 
perturbations that cleanly leads to new mechanically stable configurations to generate a suitable Markov chain to sample these states. Hence, we have been unable to use the machinery of computational statistical mechanics until now. 

Because of the above, Edwards's formalism \cite{Edwards19891080} has only been tested indirectly in most cases without solving a particular model to contrast with experimental data. By ``solving'' we mean the counting (or flat sampling) of mechanically stable states without resorting to computing the full dynamics of an experimental protocol by a molecular dynamic-type simulation. In general, these tests check for self-consistency of the theory \cite{mcnamara2009measurement}. If simple statistical mechanic simulations, like Monte Carlo simulations, were possible for large granular systems, the validity of the formalism and its limitations would had been agreed upon long ago. Interestingly, the desired Markov chain can be constructed in the pure force ensemble (for a fixed position of the grains) in two-dimensional packings \cite{tighe2008entropy}. However, this possibility has been absent in the volume ensemble so far. An interesting singular system in this respect is the 
quasi-one-dimensional model of Bowles--Ashwin (BA) \cite{PhysRevE.83.031302}. This system has an analytic solution for the entropy and has been recently compared against discrete element method (DEM) simulations of tapping; the outcome being that the flat measure used to define the entropy fails in some ranges of packing fraction \cite{1742-5468-2013-12-P12012}.   

\begin{figure}[t]
\begin{center}
    \includegraphics[width=0.4\textwidth]{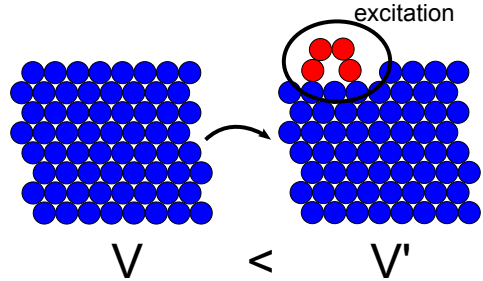}
\end{center}
    \caption{(Color online) Pictorial representation of a volume excitation created by the formation of a four-particle arch.}
    \label{excitation}
\end{figure}

The degree of complexity of a system can be largely reduced by describing it in terms of its excitations \cite{pines1999elementary}. {\it Excitations} are energy configurations that are excited from the ground state. Examples of these are the electron--hole pairs in semiconductors \cite{sze2006physics} and the magnetic monopoles \cite{castelnovo2008mono}. The natural candidates for volume excitations in static granular systems are the arches formed by the grains (see Fig. \ref{excitation}) \cite{pugnaloni2001multi,arevalo2006identification,mehta2009heterogeneities}. Arches are multiparticle structures that are stable {\it per se}. Particles in an arch support each other in the sense that, all other grains in the system being fixed, no particle of the arch can be removed without destabilizing the others. All particles in any mechanically stable granular configuration can be separated into distinct arches with each particle belonging to one and only one arch \cite{pugnaloni2001multi}. In the past, there have 
been 
proposals of using arches to describe the properties of static granular systems \cite{mehta1991vibrated} and they have shown to be important in the interpretation of several experiments \cite{PhysRevLett.86.71,pugnaloni2008,carlevaro2012arches}. However, arches have not been used as a way to define and explore static configurations in an ensemble theory.

In this work, we present an alternative approach to count mechanically stable states relying on the concept of arch. We will replace the traditional description of the geometric aspect of a microstate (i.e., the particles positions) by a description based on the properties of the arches present in the configuration. This framework provides a real opportunity to obtain the properties of model granular systems under gravity from ensemble theory, boosting our ability to test theory with experiments. Although the scheme is general, we present it here via the simplest realization of an arch system: the non-interacting rigid arch (NIRA) model. This model can be thought of as an ``ideal gas'' of arches with a single internal degree of freedom (DoF): the arch size (in number of grains). Arches do not interact directly; however, their sizes are constrained by other arches since the total number of grains in the system has to add to $N$. More sophisticated models may include further internal DoF to consider 
details such as arch shape 
and orientation, as well as arch--arch interactions. This very first model will already provide valuable insights regarding the link between physical bounding of the arch sizes and the extensivity of entropy, the effect 
that confinement 
has on the predicted volume fluctuations, and the suitability of the flat measure postulate of Edwards for the volume ensemble.

\section{Arch-based microstates} 

In the standard approach to calculating the density of states of static granular models, in the so-called volume ensemble, a volume function $W$ is used as the analogue of the Hamiltonian. Moreover, the sum over all configurations is masked with a function $Q$ that only allows mechanically stable configurations \cite{Edwards19891080}.  Unfortunately, very few configurations, out of all possible set of positions for $N$ grains, comply with mechanical stability. This has posed an important hindrance in providing even approximate solutions to the simplest models. Here, we will only consider particle positions that warrant some basic degree of mechanical stability from the start. This is done by considering arches as the basic entities (or excitations) of the system. 

Operationally, arches are defined via mutually stabilizing contacts \cite{pugnaloni2001multi}. Two grains A and B are said to be mutually stable if A supports B and B supports A. A grain ``supports'' another grain if the contact interaction is one of the necessary contacts to keep the second grain stable against gravity (there are in general a minimum of three such stabilizing contacts for each grain in three dimensions). Then, an arch is a set of connected mutually 
stabilizing grains. Since any set of stable grains can be split into disjoint subset of arches by this procedure, if $n_i$ is the number of arches consisting of $i$ grains, the following basic condition holds

\begin{equation}
\sum_{i=1}^{N} i \cdot n_i = N. \label{suma-arcos}
\end{equation} 
Here $i=1$ represents the grains that do not form an arch with any other particle in the system (i.e., none of its contacts is a mutually stabilizing contact). We will call these grains ``arches'' of size $1$.

Since each arch is a set of mutually stable grains by definition, the problem of counting all possible static configurations can be shifted to counting all possible arch configurations compatible with the given external constraints. In this respect, there are two separated sets of DoF apart from the number of particles within an arch: (i) the arch--arch interactions (i.e., all ways of ``piling'' a set of given arches in the space), and (ii) the intra-arch configurations (i.e., all stable ways of arranging a set of grains to build an arch). 

\begin{figure}[t]
\begin{center}
    \includegraphics[width=0.3\textwidth]{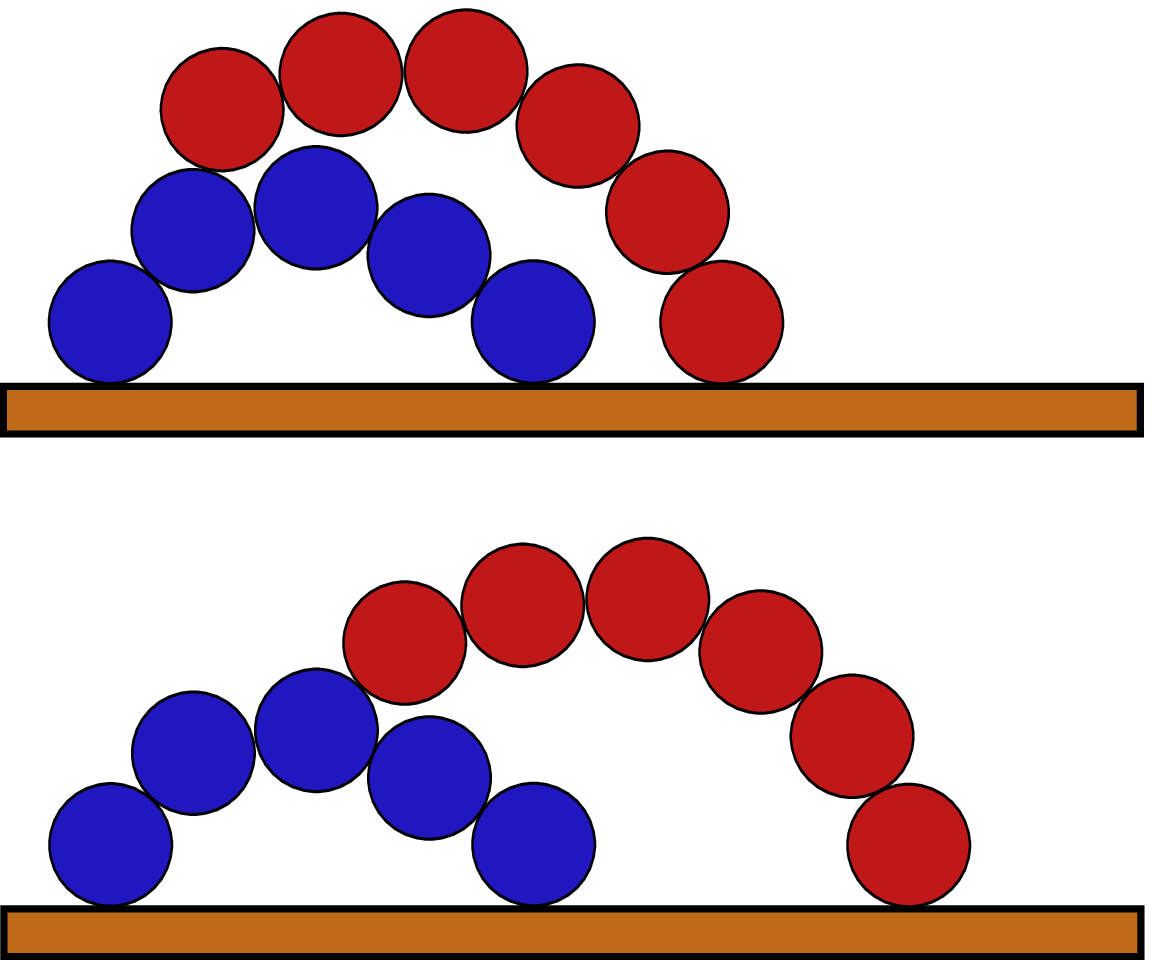}(a)
    \includegraphics[width=0.3\textwidth]{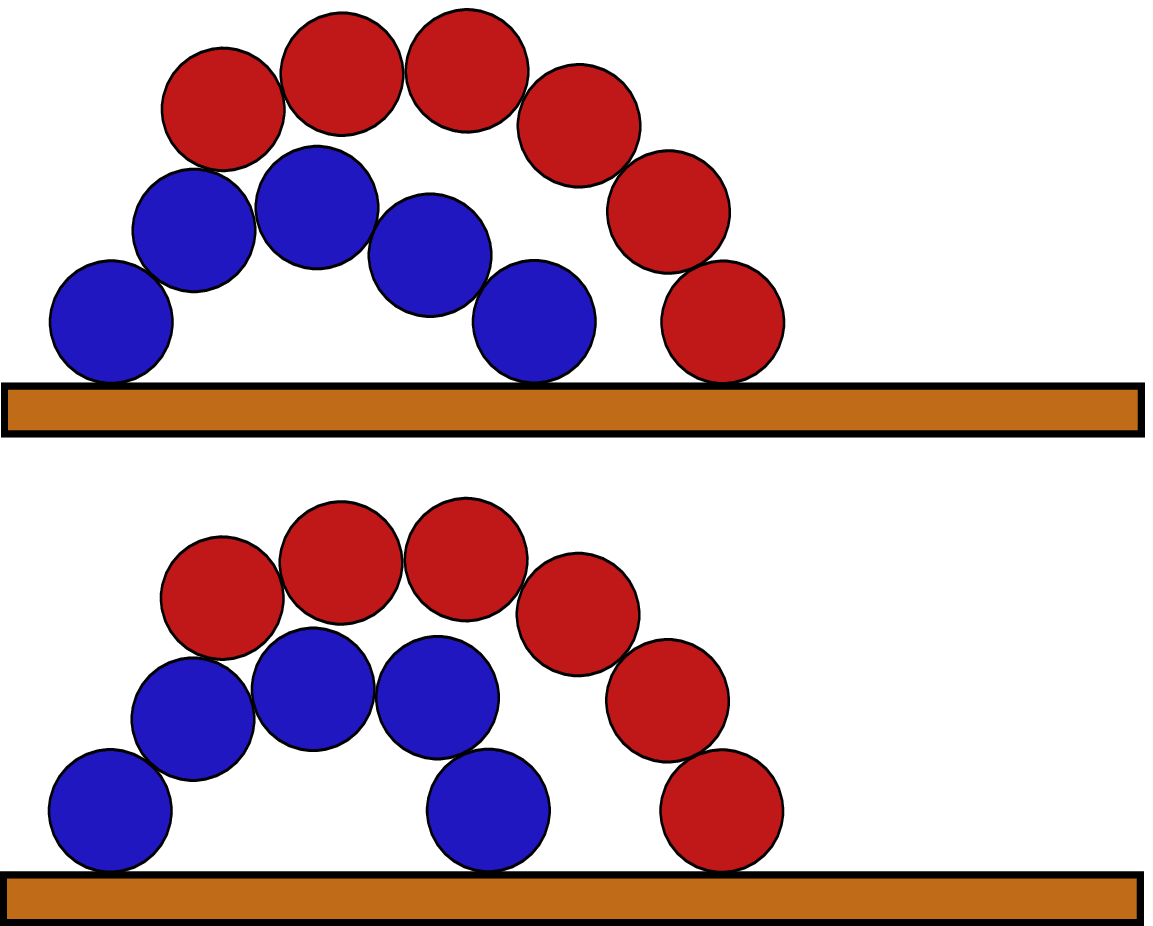}(b)    
\end{center}
    \caption{(Color online) Pictorial representation of arch configurations. (a) Two arches arranged in different configurations while keeping their shape (only a rotation and displacement separates the two mechanically stable configurations) which corresponds to what we call the arch--arch DoF. (b) Two configurations where only the shape of an arch has been changed, which corresponds to the intra-arch DoF.}
    \label{dof-arches}
\end{figure}

Figure \ref{dof-arches}(a) shows a pictorial representation of two different configurations compatible with a given set of arches with fixed size and shape (the arch--arch DoF). These variations in the way arches are locked with each other are difficult to take into account in the enumeration of states. Figure \ref{dof-arches}(b) is an example of two states where only the internal structure of one arch is different (the intra-arch DoF). Of course, both types of DoF are coupled since the arch--arch arrangements depend on the size and shape of the individual arches. In what follows, we will show how to construct the density of states (DoS) for a simple model of NIRA where (i) and (ii) are neglected. Although the model is an extreme simplification, we will learn some basic physics regarding the DoS for a static granular assembly and will show that in the high packing fraction limit results are consistent with realistic simulations of tapped granular packs. Moreover, we will discuss how more detailed DoF can be 
included.  

We split the task for the calculation of the entropy in five general steps. In the following sections we will implement these for our NIRA model as an example, bearing in mind that these steps can also be followed for more complex models. The five steps are

\subsection{Step (a). Define the microstate of the system in terms of arches} 

This initial step is strongly dependent on the DoF that will be accounted for in the particular model of interest. Instead of defining the microstate in terms of the positions of each grain, we have to define the properties of each arch. These properties are size (in number of gains), shape, position, orientation, etc. Of course, any of these properties must take values that are compatible with an stable arch (this will be care of in the next step). The arch-based description of the microstate allows the introduction of various levels of simplification while keeping the basic ingredient of mechanical stability within each arch. In section \ref{micro-NIRA} we will present the simplest representation of a configuration in this arch-based framework.

\subsection{Step (b). Define the external constraints imposed to arches.} 

Arches cannot take any shape, size or orientation if they ought to be stable under gravity. For example, in real systems, the maximum number of grains in an arch is usually bounded (e.g., due to the size of the container). In our NIRA model we will simply constraint the maximum number of particles in an arch. In section \ref{constraints} we will describe how this simple constraint is the origin of the extensivity of the entropy.

\subsection{Step (c). Define a volume function that yields the total volume of the microstate.} 

Given the microstate, one has to associate a volume to it. This volume has to be calculated in terms of the aches and their properties as defined in the microstate. The volume function can be formulated with different degrees of simplifications. We will show a simple implementation for the NIRA model in section \ref{volume-sec}.

\subsection{Step (d). Define an algorithm to sample microstates}

This step consists in generating all microstates (or sampling uniformly to meet the equal {\it a priori} postulate) defined in step (a) that comply with the constraints of step (b). In the Appendix we describe the algorithm we have implemented to generate all possible states for the NIRA model. This is possible for this simplified representation thanks to the relatively small number of states. However, for more sophisticated arch-based models, sampling will become unavoidable. The algorithm we use will serve in the future to test the goodness of sampling techniques by comparing with this exact counting.

\subsection{Step (e). Calculate the volume of each microstate generated in step (d) using the function in step (c) and build the DoS.} 

To this end we simply need to add $1$ to a volume histogram each time a configuration generated in step (d) is found to be in the given volume bin. As a result, we have the exact entropy for the $N$-particle system using equal {\it a priori} probabilities for the states since each distinguishable configuration is counted once. 

\section{The microstate in the NIRA model} 
\label{micro-NIRA}
In the NIRA model, each microstate is simply described by a vector $n$ of $N$ coordinates, $\{n_i\}=(n_{1},n_{2},...,n_{N})$ which indicates the number $n_i$ of arches consisting of $i$ particles in the system. In general, most $n_i$ will be zero since Eq. (\ref{suma-arcos}) must hold. Table \ref{tab:microstates} shows all possible configurations for a system of $8$ particles. We will not consider here DoF such as the actual shape of the arches, their positions and orientations, or how they rest on top of each other. This is, of course, a strong simplification. Although arches are not warranted to be stable against each other, their constituent grains are mutually stable by definition. This can be thought of as a non-interacting system of arches with a single internal DoF: the number of grains in the arch. 

Although the actual placement of each arch is not accounted for in the NIRA model, we do take due care of the arch permutations. Since the state is simply defined by the size of each arch, the interchange of arches of the same size leads to indistinguishable configurations. However, the interchange of arches of different sizes leads to distinguishable microstates; therefore, the associated number of distinguishable permutations of arches for a given vector $\{n_i\}$ has to be included in the DoS. If $N_A=\sum_{i=1}^N n_i$ is the total number of arches in the configuration $\{n_i\}$, then there are $N_A!/(n_{1}!...n_{N}!)$ permutations of these arches, where permutations of indistinguishable arches of the same size have been removed. 

\begin{table}
    \centering
        \begin{tabular}{ c c c c c c c c c c }
            \hline \hline
            $n_1$&$n_2$&$n_3$&$n_4$&$n_5$&$n_6$&$n_7$&$n_8$&$V\{n_i\}/N$&$N_A!/(n_{1}!...n_{N}!)$\\
            \hline
            8&0&0&0&0&0&0&0&	0.87	&	1	\\
            6&1&0&0&0&0&0&0&	0.88	&	7	\\
            5&0&1&0&0&0&0&0&	0.95	&	6	\\
            4&2&0&0&0&0&0&0&	0.89	&	15	\\
            4&0&0&1&0&0&0&0&	1.04	&	5	\\
            3&1&1&0&0&0&0&0&	0.96	&	20	\\
            3&0&0&0&1&0&0&0&	1.17	&	4	\\
            2&3&0&0&0&0&0&0&	0.90	&	10	\\
            2&1&0&1&0&0&0&0&	1.05	&	12	\\
            2&0&2&0&0&0&0&0&	1.02	&	6	\\
            2&0&0&0&0&1&0&0&	1.34	&	6	\\
            1&2&1&0&0&0&0&0&	0.97	&	12	\\
            1&1&0&0&1&0&0&0&	1.18	&	6	\\
            1&0&1&1&0&0&0&0&	1.12	&	6	\\
            1&0&0&0&0&0&1&0&	1.55	&	2	\\
            0&4&0&0&0&0&0&0&	0.91	&	1	\\
            0&2&0&1&0&0&0&0&	1.06	&	3	\\
            0&1&2&0&0&0&0&0&	1.04	&	3	\\
            0&1&0&0&0&1&0&0&	1.35	&	2	\\
            0&0&1&0&1&0&0&0&	1.25	&	2	\\
            0&0&0&2&0&0&0&0&	1.21	&	1	\\
            0&0&0&0&0&0&0&1&	1.81	&	1	\\ 
            \hline \hline
        \end{tabular}
    \caption{All microstates for an eight-particle system described by $\{n_i\}$. The second to last column is the volume per particle associated to each state using the volume function of Eqs. (\ref{totvol}) and (\ref{volume2}). The last column indicates the number of distinguishable configurations associated to each microstate $N_A!/(n_{1}!...n_{N}!)$.}
    \label{tab:microstates}
\end{table}

\section{Volume function}
\label{volume-sec}

\begin{figure}[t]
\begin{center}
    \includegraphics[width=0.3\textwidth]{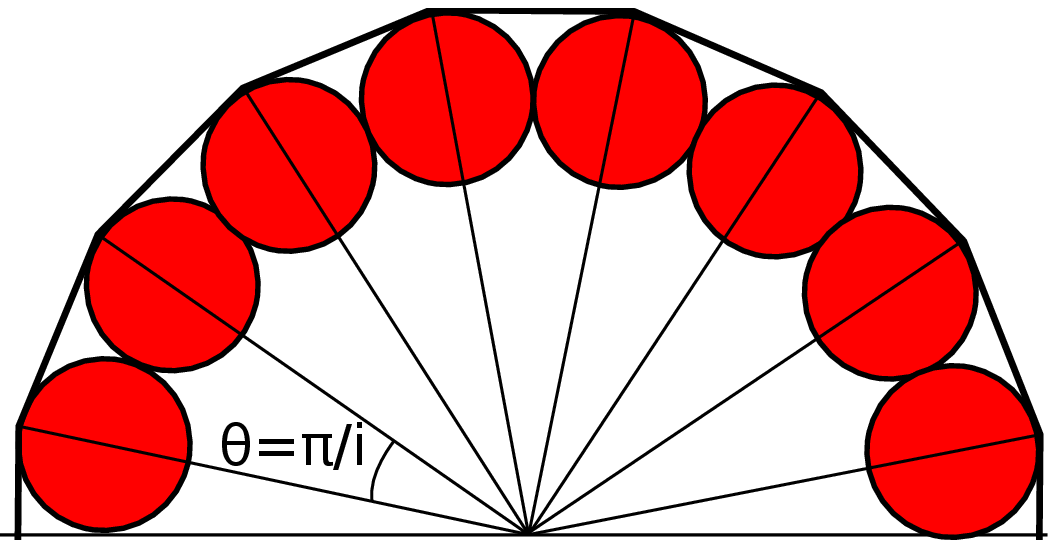} (a)
    \includegraphics[width=0.44\textwidth]{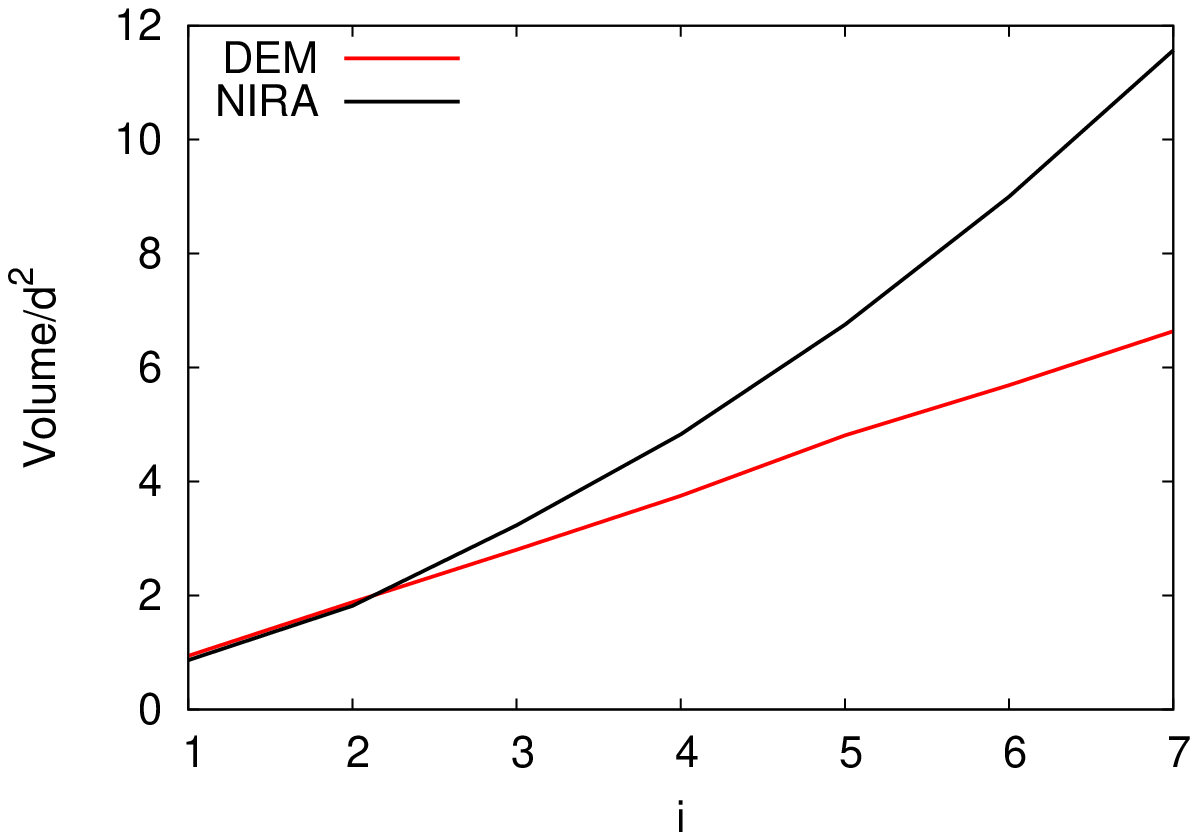} (b)
\end{center}
    \caption{(a) Example of the volume associated to an eight-particle ``rigid'' arch. (b) Comparison between the volume of an arch as a function of the arch size $i$ given by Eqs. (\ref{totvol}) and (\ref{volume2}) (black solid line) and the mean volume of arches observed in the DEM simulations described in Section \ref{dem} (red solid line). The volume of an arch in the simulation is taken as the sum of the Voronoi area of the cell of each particle in the arch. All arches of a given size are averaged over 500 configurations for a tapped system that presents a mean packing fraction $\phi=0.838$.}
    \label{vollet}
\end{figure}

We will consider as a rough approximation that the total volume of the system is the sum of the individual volumes of the arches. Since the simple microstates considered in in the NIRA model are described solely by the number of arches $\{n_i\}$ of each possible size $i$, the volume $V$ of a particular configuration $\{n_i\}$ is

\begin{equation}
V[\{n_i\}] = \sum_{i=1}^{N}{v_i n_i}, \label{totvol}
\end{equation}
where $v_i$ is the volume of an arch of $i$ disks. 

We have written a simple volume function for each arch having in mind a 2D system where grains are represented as equal-sized disks of diameter $d$. The volume $v_i$ assigned to any arch consisting of $i$ particles with $i>2$ is given by

\begin{equation}
v_i = i\left( \frac{d}{2} \right)^2 \tan\left( \frac{\pi}{2 i} \right) \left[ 1 + \left(\tan\left( \frac{\pi}{2 i} \right)\right)^{-1} \right]^2 \mathrm{\ for\ }i>2. \label{volume2}
\end{equation}
This volume corresponds to half the area of the regular polygon that inscribes all disks in a semi-circular arch (see Fig. \ref{vollet}(a)). These arches are rigid in the sense that deviations from semi-circular array of particles are forbidden. In the limit when $N$ is large, the volume increases quadratically with $i$: $V\sim (i d)^2/(2\pi)$. 

For arches of size $1$, we have used $v_1=\frac{\sqrt{3}}{2} d^2$. This volume corresponds to the area of the hexagon that inscribes a disk consistent with the disk being in a triangular ordered environment. For the 2D case we consider here, this is a fair approximation since mono-sized disks tend to order into the closest pack density, particularly if they do not form arches.

Finally, for arches of two particles, where a polygon cannot be defined, we assigned $v_2 = \alpha v_1$. The factor $\alpha$ is taken to be slightly larger than $2$ so as to assign to the two-particle arch a volume greater than the volume occupied by two separate grains forming arches of size $1$. The actual value of this factor can change the range of volumes the entire system can achieve since arches of size two are the most common arches. We found that $\alpha = 2.1$ yields volumes $V$ in a range similar to the ones observed in the DEM simulations. 

Figure \ref{vollet}(b) shows the volume proposed as a function of the arch size $i$. For comparison, the mean volume of arches detected in a DEM simulation (see Section \ref{dem}) is also shown. The volume of each arch in the simulations is taken as the sum of the area of the Voronoi cell of each disk in the arch. It is clear that our volume function overestimates the volume of large arches. Notice that the proposed volume function could be in principle replaced by this ``empirical'' information of the mean volume of the arches. However, we have to bear in mind that this ``empirical'' data has included all possible arch-arch DoF since the volume of each arch is affected by the neighbor grains in the pack observed in the simulations. Doing such replacement of the volume function would not allow to separate the effects of different DoF.

The complexity of the volume function will depend on the degree of detail of the intra-arch and the arch--arch DoF included in the definition of the microstate. Even within the NIRA model, a similar definition can be used to consider 3D systems by assigning a meaningful 3D volume to arches of different sizes. We have considered 2D systems since we have available DEM simulation data to compare with. 

\section{External constraints and entropy extensivity} 
\label{constraints}

We have found that the entropy of a static granular system is, in principle, non-extensive if all possible microstates are taken into account. Consider two isolated identical $N$-particle systems that at a given volume $V$ have $\mathcal{N}$ possible configurations each. The number of configurations associated to the combined $2N$-particle system at volume $2V$, if correlations are neglected, should be $\mathcal{N}^2$ (leading to doubling the entropy). However, in principle, the combined system can now form arches larger than $N$ particles that were unavailable for the separate $N$-particle systems. These new states represent a finite portion of the entropy. As a consequence, there exist a large number of new configurations available to the combined system not reachable in the isolated systems, which results in a non-additive entropy. Although for small $N$ correlations in granular samples do lead to non-extensive properties \cite{lechenault2006free}, one should expect that correlations can be 
neglected in larger systems and all macroscopic 
properties should be extensive.

\begin{figure}[t]
\begin{center}
    \includegraphics[width=0.6\textwidth]{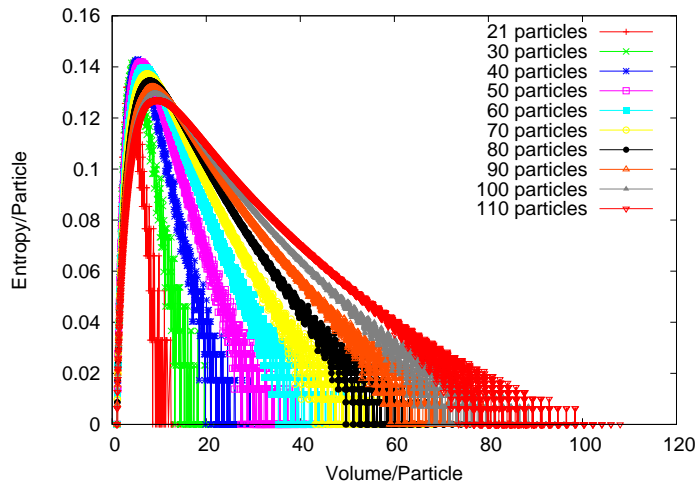}
\end{center}
    \caption{Entropy per particle as a function of volume per particle for different system sizes without a cut-off in the arch size. The entropy is clearly non-additive, due to long range correlations.}
    \label{non-extensive-entropy}
\end{figure}

Fig. \ref{non-extensive-entropy} shows the entropy we calculated for the 2D NIRA model per unit particle as a function of the specific volume ($V/N$) for systems of 
different sizes $N$. As we can see, $S/N$ for a given $V/N$ depends on the system size, indicating that $S$ is not a homogeneous function of $V$ and $N$.

This non-additive nature of the entropy lies on the unphysical assumption that arches of all sizes can be present in a system; implying that correlations can span the system. In real systems, the maximum number of grains in an arch is usually bounded (e.g., due to the size of the container). In order to capture this feature, we restrict the maximum size $L$ that an arch can be by simply adding the constraint $n_i = 0 \  \forall  i>L$. This constraint cuts off correlations, which leads to an additive entropy consistent with standard thermodynamics (see section \ref{results}).

\section{Validation} 

\begin{figure}[t]
\begin{center}
    \includegraphics[width=0.6\textwidth]{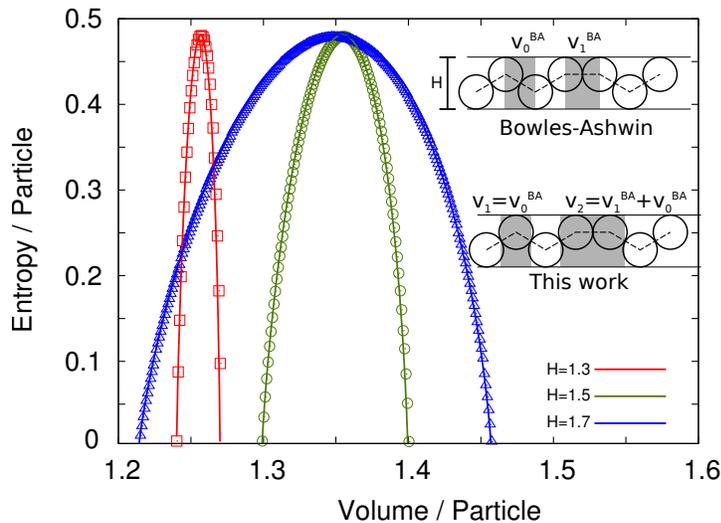}
\end{center}
    \caption{(Color online) The entropy for the quasi-1D BA model for different widths, $H$, of the system. The solid lines correspond to the analytic result in the thermodynamic limit and symbols to our exact calculation for 1000 particles at $H=1.3$ (red squares), 1.5 (green circles) and 1.7 (blue triangles). The values for the basic volumes in the BA model $v_1^{BA}$ and $v_2^{BA}$ depend on the width of the container (see upper inset). Since we focus on the volume of arches and BA focus on the volume associated to branch vectors, in our arch representation $v_1$ is equivalent to $v_0^{BA}$, and $v_2$ corresponds to $v_1^{BA}+v_0^{BA}$ (see lower inset) \cite{PhysRevE.83.031302}.}
    \label{ba-model}
\end{figure}

In the limit where the maximum number of particles in arches is restricted to two ($L=2$), the NIRA model becomes a realization of the BA model. This is a quasi-1D system consisting of disks of diameter $d$ confined between two walls separated by less than $(1+\sqrt{3/4}) d$. Each disks can be mechanically stable only if it has three contacts (not all on the same semi-circle). Under these conditions each disk needs to be in contact with one of the walls. Disks can be arranged along the confined quasi-1D space in a sequence by touching the previous disk and one of the walls (see inset of Fig. \ref{ba-model}). However, no more than two consecutive disks can be aligned with the same wall for all of them to be stable with three contacts not in the same semi-circle. This leads to a simple analytic solution since all possible configurations correspond to the number of all possible consecutive aligned pair of disks  \cite{PhysRevE.83.031302}. Two consecutive disks aligned on a wall constitute an arch of size 
two in our 
representation. Since larger sets of this type are not stable in the BA model, the natural cutoff for this system becomes $L=2$.   

We have plotted in Fig. \ref{ba-model} the entropy for the BA model along with our exact calculation for $10^3$ grains. As it is to be expected, the two independent calculations agree. This validates our general technique to generate the microstates.


\section{Results for a 2D column}
\label{results}

\begin{figure}[t]
\begin{center}
    \includegraphics[width=0.6\textwidth]{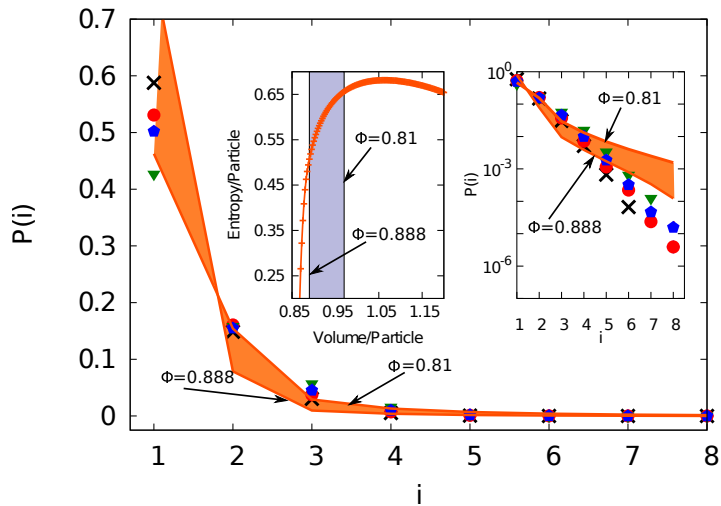}
\end{center}
    \caption{(Color online) The arch size distribution ($P(i)=n_i/N$) predicted by the NIRA model (shaded area) for $400$ grains setting $L=8$ along with the results from DEM simulations (symbols) of tapped 2D system ($512$ frictional disks in a box 13-particle diameters wide \cite{arevalo2006identification}). Each symbol corresponds to DEM results for a given tap amplitude that yields a particular mean volume per particle: 0.884 (black crosses, $\phi=0.888$), 0.905 (red circles, $\phi=0.867$), 0.929 (blue pentagons, $\phi=0.845$), 0.97 (green down triangles, $\phi=0.810$). The shaded area covers the same range of volumes in the model. Left inset: entropy per particle as a function of volume (the shaded area indicates the range of mean volumes considered in the main plot). Right inset: Same as main figure but in semi-log scale.}
    \label{rlu}
\end{figure}

Using the NIRA model, we have computed the entropy as a function of the volume, the volume fluctuations and the arch size distribution $n_i$ for the 2D realization we consider here. 

The arch size distribution $n_i$ is known by construction for each configuration. Hence, for a given volume bin, we can average $n_i$ over all corresponding configurations. Based on DEM data of arches identified in 2D columns \cite{arevalo2006identification} against which we would like to compare these results of the NIRA model in the next section, we have set $L=8$ in accordance with the larger arch observed in the simulations for the system of reference (512 frictional monosized disks in a box of width $12.39 d$ with $d$ the disk diameter). It has been observed that $n_i$ falls very rapidly and even for wide systems the larger arch found is not much larger (10 disks in a system $24.78 d$-wide) \cite{carlevaro2012arches}.   

In Fig. \ref{rlu} we show $n_i$ corresponding to the lower ($V/N\approx0.884\rightarrow\phi\approx0.888$) and upper ($V/N\approx0.97\rightarrow\phi\approx0.810$) limits of a range of volumes indicated in the left inset, where the calculated entropy is displayed.  The corresponding $n_i$ curves are indicated in orange in the main plot and the area between the two has been shaded. This region of volumes has been chosen since it coincides with the range of packing fractions that have been obtained in DEM simulations (see symbols and discussion in section \ref{dem}). We can see that $n_i$ decays rapidly, with large arches being less likely for a given volume of the system. In accordance with intuition, for larger volumes there is a higher incidence of large arches (for clarity, see log plot in the right inset in Fig. \ref{rlu}).

\begin{figure}
\begin{center}
    \includegraphics[width=0.4\textwidth]{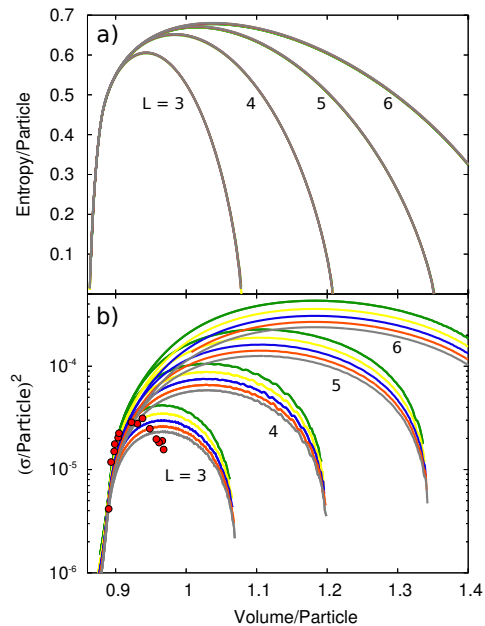}
\end{center}
    \caption{(Color online) (a) Entropy per unit particle as a function of unit volume for the NIRA model using $N=500$ (green), $600$ (yellow), $700$ (blue), $800$ (orange) and to $900$ (gray) (curves are indistinguishable in this scale) and for different cutoff $L$ (see labels). (b) Volume fluctuations as a function of volume per unit particle from numerical differentiation of part (a). The red symbols correspond to the DEM simulation of tapped disks \cite{pugnaloni2010towards}.}
    \label{flu}
\end{figure}

Figure \ref{flu}(a) shows the entropy per unit particle $S(V)/N$ obtained for our model using various $L$ and $N$ ($2\leq L\leq 6$ and $300\leq N\leq 1000$). This converges rapidly as we increase $N$ beyond $500$, which indicates that the entropy complies with being additive for large systems. The DoS presents a maximum, as in all models of static granular systems, since the maximum possible volume is bounded and this leads to inversion population of the states. As $L$ increases, larger volumes are possible and the DoS grows at the maximum, as it is expected. Interestingly, for small volumes, the entropy per unit particle is independent of the cut-off $L$. This is due to the fact that small volumes correspond to configurations where most particles form arches of size 1 and only a few arches of two or more particles are found. Therefore, the imposed cut-off does not limit the number of configurations compatible with the given volume.  

We have also calculated the volume fluctuations characterized by the variance $\sigma_V^2$ of the volume as $\sigma_V^2=\lambda\chi^2\partial V/\partial \chi$. Here, $\chi$ is the compactivity defined as the intensive variable conjugate to $V$, i.e., $\chi^{-1}=\partial S/\partial V$; whereas $\lambda$ is the equivalent to the Boltzmann constant that we set to $1$. Figure \ref{flu}(b) displays $\sigma_V^2$ obtained after numerical differentiation of $S(V)$ in Fig. \ref{flu}(a). As we can see, a maximum in the fluctuations is observed. This maximum in the fluctuations has been observed before in simulations and experiments \cite{pugnaloni2010towards,1742-5468-2013-12-P12012}. The symbols in Fig. \ref{flu}(b) correspond to DEM simulations and will be discussed in the next section. Here, as for the entropy, fluctuations are insensitive to the arch size cut-off for small system volumes. However, for larger volumes, fluctuations are predicted to increase with increasing $L$. This suggest that systems with a 
tendency to form larger arches (for example due to a large static friction in the grain--grain interaction) should display enhanced volume fluctuations. 

\section{Comparison with DEM}
\label{dem}
In order to compare the predictions with realistic simulations of granular packs, we have carried out DEM simulations of the tapping of a 2D column of frictional disks. $N=512$ soft disks, interacting through linear spring and dash-pot forces in both normal and tangential directions, are placed in a rectangular box. In the tangential direction, the Coulomb criterion is used to switch between static and dynamic friction. Tapping is simulated by moving the confining box following a harmonic pulse of given amplitude and duration. Details of the simulation can be found in Refs. \cite{arevalo2006identification,pugnaloni2011master}. The tapping protocol leads to states amenable of statistical description using equilibrium ensembles \cite{pugnaloni2010towards}. In particular, after a transient, using a given tap amplitude, we sample configurations with well defined mean volume and volume fluctuations. In our DEM simulations, we are also able to identify arches and build $n_i$. These arches are identified by 
following the history of all contacts and defining the first two contacts formed by a particle that can support its wright based on geometrical considerations. Then, the list of mutually stabilizing contacts and the arches can be extracted. For a detailed description of the algorithm see Ref. \cite{arevalo2006identification}.

For a given tap intensity, we tap the system of disks until we reach the steady state. In the steady state the packing fraction and stress tensor fluctuate around well defined values. In the steady state we tap the system 500 times and average volume and arch size distributions. The volume fluctuations obtained as the standard deviation of the volume distribution over the steady state is averaged over 20 independent repetitions of the simulation. 

As shown in Fig. \ref{rlu}, $n_i$ for $L=8$ for the NIRA model captures rather well the distribution found in our simulations for small arches. However, the incidence of large arches is overestimated by the NIRA model (see right inset in Fig. \ref{rlu}). This is against expectations. Equation (\ref{volume2}) overestimates the real volume of arches. Most arches are flatter in real systems (these intra-arch DoF are not included) and the real volume they occupy may be reduced by the presence of another arch filling the cavity underneath (this corresponds to the neglected arch--arch DoF). Hence, to comply with a given volume, one should expect that the NIRA model will count configurations with few large arches. The fact that a higher number of large arches (instead of the expected lower number) is predicted by the NIRA model than observed in the simulations indicates that there must be a different origin for this bias. The other approximation introduced is the assumption of a flat measure in the Edwards ensemble.
 It seems that this equal {\it a priori} probability of the states is unsuitable and bias the statistics on arches to the point that the expected underestimation of large arches of the NIRA model is reverted. According to this observations, configurations having large arches seem to require a lower weight than microstates having smaller arches. Previous studies have also shown that the equal {\it a priori} postulate may not be suitable in describing steady states of tapped packs in very small systems \cite{gao2009experimental,1742-5468-2013-12-P12012}.

Figure \ref{flu}(b) compares the fluctuation $\sigma_V^2$ from the DEM simulations with those predicted by the NIRA model. As observed in experiments and simulations \cite{pugnaloni2010towards}, and also found in the BA model \cite{1742-5468-2013-12-P12012}, fluctuations present a maximum at a volume lower than the volume where the maximum of entropy is located. Considering the strong simplifications, the qualitative agreement between the NIRA model and DEM results is fair for low volumes where the entropy is independent of the arch size cut-off $L$. Notice that, since the NIRA model does not consider arch--arch interactions, macrostates compatible with the presence of few isolated arches should be better represented. Indeed, the low volume macrostates contain few arches and correspondingly the NIRA model, irrespective of $L$, is a good approximation that results in reasonable predictions for the volume fluctuations in this limit.

\section{Conclusions} 

The arch-based statistics ensures a basic mechanical stability of the grains in any proposed configuration of a model system. We have established five basic steps to calculate the entropy of any arch-based model of grains under gravity. In particular, we have generated all configurations for a non-interacting rigid arch model for systems of up to thousand particles. 

The arch size distribution and volume fluctuations predicted are in overall agreement with realistic simulations of tapped 2D systems, particularly at low system volumes where arch--arch interactions are less prominent. Apart from the various simplifications of the model, deviations from the DEM results may be due in part to the unsuitability of the Edwards flat measure. We found evidence that configurations containing large arches are overrepresented by using the equal {\it a priori} probability postulate. However, we have to bear in mind that the relative number of microstates of a given volume bin is affected by the neglected number of different contact force arrangements compatible with each of the geometric microstate in the bin. This may be also a cause for the deviations of the predicted arch size distribution when compared with the DEM results. Moreover, we showed that fluctuations can be affected by external constraints since they are predicted to grow if larger arches are allowed. Finally,
 we showed that an additive entropy is found only if there exists a physical cutoff to the largest possible arch in the assembly of grains. 

One can make important improvements in the quantitative predictions by sophisticating the microstate definition and selecting more accurate volume functions in the arch model. A natural extension is to use a more ``flexible'' description of the arch shape as proposed in Ref. \cite{PhysRevLett.86.71}. However, due to the continuous nature of this extra DoF, this leads to a cumbersome computational task if all possible states shall be generated as we did in the present work. Fortunately, the arch-based statistics is amenable of Monte Carlo sampling. Notice that simple rules to generate a static configuration from an existing one can be given (e.g., by removing a grain from one arch and inserting it in a different arch) to generate a Markov chain \cite{demian}. This gives an opportunity for a range of more complex arch-based models to be studied via ensemble theory and the results contrasted against experiments. An interesting reference case for the validation of a flexible arch model would be 
the extension of the BA model to slightly wider systems for which analytical approximations are available \cite{ashwin2009complete}. 

\ack
We acknowledge fruitful discussions with P. A. Gago.

\section*{Appendix A: State counting in the NIRA model}
\label{appendix}

Since the definition of the configuration in the NIRA model via $n_i$ is simple, we have developed an algorithm to generate all possible configurations (see section 4). In brief, we create all possible vectors $n_i$ compatible with Eq. (\ref{suma-arcos}). Since we have to leave aside all configurations in which $n_i\neq 0$ for $i>L$, the number of configurations is greatly reduced. We can generate all configurations for systems of up to thousand particles in minutes (for $L<4$) or in a few hours (for $5<L<10$). Notice that other approaches to count or sample mechanically stable states are able to consider in a natural way more complex granular models. However, the application of these approaches has been limited to very small systems \cite{C2SM06898B,gao2009experimental,hoy2012structure,PhysRevLett.101.128001,asenjo2014numerical}.   
Table \ref{tab:microstates} shows, as an example, all configurations for an eight-particle system in the NIRA model where the state is defined via $\{n_i\}=(n_{1},n_{2},...,n_{N})$. 

The overall idea of the algorithm we use to generate all possible configurations in step (d) of section 2 is the following.  First, we construct all possible configurations with $n_{1} = N$, then all possible configurations with $n_{1} = N-1$, then with $n_{1} = N-2$, etc. Note that the algorithm is iterative given that for a fixed $n_{1} = i$, all possible configurations can be ordered in the same way, beginning with the $n_{2} = \mathrm{int}((N-i)/2)$, and following with all configurations with $n_{2} = \mathrm{int}((N-i-1)/2)$, etc. Of course, valid configurations are those that comply with Eq. (\ref{suma-arcos}). A large number of configurations $\{n_i\}$ that do not comply with Eq. (\ref{suma-arcos}) can be easily avoided to reduce computation. The algorithm presented in the pseudocode of Fig. \ref{pseudocode} takes partial advantage of this. Using this algorithm, it is simple to add a constraint $L$ to cut-off the maximum arch length. To do so, whenever $c > L$, $c$ is taken back to be $L$ 
and an arch is subtracted from $c=L-1$.

\begin{figure}
\begin{alltt}
\footnotesize
n(1)=N; n(2:N)=0; c=1; ck=N
\textbf{do forever}
  \textbf{if} ( ck < N ) \textbf{then}
    n(c)++; ck=ck+c
  \textbf{else}
    \textbf{if} ( ck == N ) \textbf{REPORT VALID STATE}
    \textbf{if} ( n(c) == 1 ) \textbf{then}			
      n(c)--; ck=ck-c
      \textbf{while} ( n(c-1) == 0 ) \textbf{do}
        \textbf{if} ( c == 2 ) \textbf{then}
          \textbf{STOP}
        \textbf{else}
          c--
        \textbf{end if}			
      \textbf{end while}				
      n(c-1)--; ck=ck-(c-1)			
    \textbf{else}
      n(c)--; ck=ck-c; c++
    \textbf{end if}	
  \textbf{end if}
\textbf{end do} 
\end{alltt}
    \caption{Pseudocode for the algorithm described in the text. The variable $c$ points to a position in the vector $n$ where $\{n_i\}$ is stored. The variable $ck$ is the number of particles of the proposed configuration. Whenever $ck$ is greater than $N$, arches must be subtracted.}
    \label{pseudocode}
\end{figure}

\begin{table}
    \centering
    \begin{tabular}{ c  c  r  c }
        \hline \hline
        $N$   & $L$ & No. valid states  & Accept. ratio (\%) \\
        \hline
        100               & 7 & 596,763 & 3.53  \\
        100               & 8 & 1,527,675 & 3.56  \\
        100               & 9 & 3,314,203 & 3.55  \\
        150               & 7 & 5,326,852 & 2.78  \\
        150               & 8 & 18,352,987 & 2.85  \\
        150               & 9 & 52,393,552 & 2.88  \\
        \hline \hline
    \end{tabular}
    \caption{Performance of the algorithm used to generate all possible arch configurations for different system sizes $N$ and cutoff $L$. The last two columns show the number of microstates obtained for all possible volumes and the percentage they represent in the set of all configurations tested by the algorithm (acceptance ratio).}
    \label{tab:efficiency}
\end{table}

In Table \ref{tab:efficiency} we show the performance of our algorithm. The number of valid microstates for different $N$ and different cut-off $L$ are shown along with the percentage that these valid configurations represent in the total of configurations tested. As can be seen, the algorithm described above is quite inefficient. Further improvements can be made by avoiding adding and subtracting one arch at a time. Also, if one can predict the next microstate in the ordered set that will comply with Eq. (\ref{suma-arcos}) a maximum efficiency could be reached.

\section*{References}

\bibliographystyle{unsrt}
\bibliography{biblio.bib}

\end{document}